\input harvmac
\noblackbox
\input epsf
\def\abstract#1{
\vskip .5in\vfil\centerline
{\bf Abstract}\penalty1000
{{\smallskip\ifx\answ\bigans\leftskip 1pc \rightskip 1pc 
\else\leftskip 1pc \rightskip 1pc\fi
\noindent \abstractfont  \baselineskip=12pt
{#1} \smallskip}}
\penalty-1000}
%
\def\hth/#1#2#3#4#5#6#7{{\tt hep-th/#1#2#3#4#5#6#7}}
\def\nup#1({Nucl.\ Phys.\ $\us {B#1}$\ (}
\def\plt#1({Phys.\ Lett.\ $\us  {B#1}$\ (}
\def\cmp#1({Comm.\ Math.\ Phys.\ $\us  {#1}$\ (}
\def\prp#1({Phys.\ Rep.\ $\us  {#1}$\ (}
\def\prl#1({Phys.\ Rev.\ Lett.\ $\us  {#1}$\ (}
\def\prv#1({Phys.\ Rev.\ $\us  {#1}$\ (}
\def\mpl#1({Mod.\ Phys.\ Let.\ $\us  {A#1}$\ (}
\def\atmp#1({Adv.\ Theor.\ Math.\ Phys.\ $\us  {#1}$\ (}
\def\ijmp#1({Int.\ J.\ Mod.\ Phys.\ $\us{A#1}$\ (}
\def\jhep#1({JHEP\ $\us {#1}$\ (}

\def\subsubsec#1{\vskip 0.2cm \goodbreak \noindent {\it #1} \br}

\def\bx#1{{\bf #1}}
\def\cx#1{{\cal #1}}
\def\tx#1{{\tilde{#1}}}
\def\hx#1{{\hat{#1}}}
\def\rmx#1{{\rm #1}}
\def\us#1{\underline{#1}}
\def\fc#1#2{{#1\over #2}}
\def\frac#1#2{{#1\over #2}}

\def\br{\hfill\break}
\def\ni{\noindent}

\def\al{\alpha}\def\be{\beta}\def\ga{\gamma}
\def\p{\partial}

\def\CY{Calabi--Yau\ }
\def\vx#1{{\vec#1}}

\def\IP{\bx P}\def\CC{\bx C}
\def\lm#1{l^{(#1)}}
\def\ss{\scriptstyle}

\def\omm#1{#1^{(o)}}\def\cmm#1{#1^{(c)}}
\def\Wtd{\cx W_{D=2}}
\def\OM#1{\Omega^{(#1,0)}}
\def\zo{z_0}\def\ty{\tilde{y}}
\def\yyy{ }
\lref\kbmv{S.~Katz, P.~Mayr and C.~Vafa,
``Mirror symmetry and exact solution of 4D N = 2 gauge theories. I,''
Adv.\ Theor.\ Math.\ Phys.\  {\bf 1}, 53 (1998), hep-th/9706110;\br
P.~Berglund and P.~Mayr,
``Heterotic string/F-theory duality from mirror symmetry,''
Adv.\ Theor.\ Math.\ Phys.\  {\bf 2}, 1307 (1999), hep-th/9811217.}
\lref\agm{P.~S.~Aspinwall, B.~R.~Greene and D.~R.~Morrison,
``Measuring small distances in N=2 sigma models,''
Nucl.\ Phys.\ B {\bf 420}, 184 (1994), hep-th/9311042.}
\lref\wolfg{W.~Lerche,
``Fayet-Iliopoulos potentials from four-folds,'' JHEP {\bf 9711}, 004 (1997),
hep-th/9709146.}
\lref\jan{
M.~Haack and J.~Louis,
``M-theory compactified on Calabi-Yau fourfolds with background flux,''
Phys.\ Lett.\ B {\bf 507}, 296 (2001), hep-th/0103068.}
\lref\wittop{
E.~Witten,
``On The Structure Of The Topological Phase Of Two-Dimensional Gravity,''
Nucl.\ Phys.\ B {\bf 340}, 281 (1990);\br
E.~Witten,
``Mirror manifolds and topological field theory,''
hep-th/9112056.}
\lref\spgeo{A.~Strominger,
``Special Geometry,''
Commun.\ Math.\ Phys.\  {\bf 133}, 163 (1990);\br
P.~Candelas and X.~de la Ossa,
``Moduli space of \CY manifolds,''
Nucl.\ Phys.\ B {\bf 355}, 455 (1991).}
\lref\tfii{
P.~Candelas, X.~C.~De La Ossa, P.~S.~Green and L.~Parkes,
``A pair of Calabi-Yau manifolds as an exactly soluble superconformal  theory,''Nucl.\ Phys.\ B {\bf 359}, 21 (1991);\br
P.~S.~Aspinwall and D.~R.~Morrison,
``Topological field theory and rational curves,''
Commun.\ Math.\ Phys.\  {\bf 151}, 245 (1993), hep-th/9110048.}
\lref\wlsm{E.~Witten,
``Phases of N = 2 theories in two dimensions,''
Nucl.\ Phys.\ B {\bf 403}, 159 (1993), hep-th/9301042.}
\lref\amv{M.~Atiyah, J.~Maldacena and C.~Vafa,
``An M-theory flop as a large n duality,'' hep-th/0011256;\br
B.~S.~Acharya,
``On realising N = 1 super Yang-Mills in M theory,'' hep-th/0011089.}
\lref\hkty{S.~Hosono, A.~Klemm, S.~Theisen and S.~T.~Yau,
``Mirror symmetry, mirror map and applications to Calabi-Yau hypersurfaces,''
Commun.\ Math.\ Phys.\  {\bf 167}, 301 (1995), hep-th/9308122.}
\lref\ksp{M.~Bershadsky, S.~Cecotti, H.~Ooguri and C.~Vafa,
``Kodaira-Spencer theory of gravity and exact results for 
quantum string amplitudes,''
Commun.\ Math.\ Phys.\  {\bf 165}, 311 (1994), hep-th/9309140.}
\lref\gvw{S.~Gukov, C.~Vafa and E.~Witten,
``CFT's from Calabi-Yau four-folds,''
Nucl.\ Phys.\ B {\bf 584}, 69 (2000), Erratum-ibid.\ B {\bf 608}, 477 (2000),
hep-th/9906070.}
\lref\cec{S.~Cecotti,
``N=2 Landau-Ginzburg versus Calabi-Yau sigma models: 
Nonperturbative aspects,''
Int.\ J.\ Mod.\ Phys.\ A {\bf 6}, 1749 (1991).}
\lref\oo{H.~Ooguri, Y.~Oz and Z.~Yin,
``D-branes on Calabi-Yau spaces and their mirrors,''
Nucl.\ Phys.\ B {\bf 477}, 407 (1996), hep-th/9606112.}
\lref\wittopop{E.~Witten,
``Chern-Simons gauge theory as a string theory,'' hep-th/9207094.}
\lref\cyper{V.~V.~Batyrev,
``Dual Polyhedra and Mirror Symmetry for Calabi-Yau Hypersurfaces 
in Toric Varieties'', alg-geom/9310003;\br
V.~V.~Batyrev and D.~van Straten,
``Generalized hypergeometric functions and rational curves on Calabi-Yau complete intersections in toric varieties,''
Commun.\ Math.\ Phys.\  {\bf 168}, 493 (1995), alg-geom/9307010.}
\lref\ov{H.~Ooguri and C.~Vafa,
``Knot invariants and topological strings,''
Nucl.\ Phys.\ B {\bf 577}, 419 (2000),hep-th/9912123.}
\lref\gmp{B.~R.~Greene, D.~R.~Morrison and M.~R.~Plesser,
``Mirror manifolds in higher dimension,''
Commun.\ Math.\ Phys.\  {\bf 173}, 559 (1995), hep-th/9402119.}
\lref\mal{J.~Maldacena,
``The large N limit of superconformal field theories and supergravity,''
Adv.\ Theor.\ Math.\ Phys.\  {\bf 2}, 231 (1998), hep-th/9711200.}
\lref\vafaln{C.~Vafa,
``Superstrings and topological strings at large N,''
hep-th/0008142.}
\lref\hori{K.~Hori,
``Linear models of supersymmetric D-branes,'', hep-th/0012179.
}
\lref\witcs{E.~Witten,
``Chern-Simons gauge theory as a string theory,''
hep-th/9207094.}
\lref\vafaext{C.~Vafa,
``Extending mirror conjecture to Calabi-Yau with bundles,''
hep-th/9804131.}
\lref\swp{
N.~Seiberg and E.~Witten,
``Electric - magnetic duality, monopole condensation, and confinement in N=2 supersymmetric Yang-Mills theory,''
Nucl.\ Phys.\ B {\bf 426}, 19 (1994),
Erratum-ibid.\ B {\bf 430}, 485 (1994), hep-th/9407087;
``Monopoles, duality and chiral symmetry breaking in N=2 supersymmetric QCD,''
Nucl.\ Phys.\ B {\bf 431}, 484 (1994), hep-th/9408099.
}
\lref\netst{
S.~Kachru, A.~Klemm, W.~Lerche, P.~Mayr and C.~Vafa,
``Nonperturbative results on the point particle limit of N=2 heterotic string compactifications,''
Nucl.\ Phys.\ B {\bf 459}, 537 (1996), hep-th/9508155;\br
A.~Klemm, W.~Lerche, P.~Mayr, C.~Vafa and N.~P.~Warner,
``Self-Dual Strings and N=2 Supersymmetric Field Theory,''
Nucl.\ Phys.\ B {\bf 477}, 746 (1996), hep-th/9604034.
}
\lref\TV{T.~R.~Taylor and C.~Vafa,
``RR flux on Calabi-Yau and partial supersymmetry breaking,''
Phys.\ Lett.\ B {\bf 474}, 130 (2000), hep-th/9912152.}
\lref\CIV{F.~Cachazo, K.~Intriligator and C.~Vafa,
``A large N duality via a geometric transition,''
Nucl.\ Phys.\ B {\bf 603}, 3 (2001), hep-th/0103067;\br
J.~D.~Edelstein, K.~Oh and R.~Tatar,
``Orientifold, geometric transition and large 
N duality for SO/Sp gauge  theories,''
JHEP {\bf 0105}, 009 (2001), hep-th/0104037;\br
F.~Cachazo, S.~Katz and C.~Vafa,
``Geometric Transitions and N=1 Quiver Theories,''
hep-th/0108120.}
\lref\ssb{
P.~Mayr,
``On supersymmetry breaking in string theory and its realization in brane  worlds,''
Nucl.\ Phys.\ B {\bf 593}, 99 (2001), hep-th/0003198.
}
\lref\ffsp{
P.~Mayr,
``Mirror symmetry, N = 1 superpotentials and tensionless strings on  Calabi-Yau four-folds,''
Nucl.\ Phys.\ B {\bf 494}, 489 (1997), hep-th/9610162.
}
\lref\HV{K.~Hori and C.~Vafa,``Mirror symmetry,''hep-th/0002222.}
\lref\HIV{K.~Hori, A.~Iqbal and C.~Vafa,``D-branes and mirror symmetry,''
hep-th/0005247.}
\lref\wip{W. Lerche and P. Mayr, work in progress.}
\lref\AViii{M.~Aganagic and C.~Vafa,
``Mirror symmetry and a G(2) flop,''
hep-th/0105225.}
\lref\AVii{M.~Aganagic, A.~Klemm and C.~Vafa,
``Disk instantons, mirror symmetry and the duality web,''
hep-th/0105045.}
\lref\AVi{M.~Aganagic and C.~Vafa,
``Mirror symmetry, D-branes and counting holomorphic discs,''
hep-th/0012041.}
\lref\Witbranes{
E.~Witten,
``Branes and the dynamics of {QCD},''
Nucl.\ Phys.\ B {\bf 507}, 658 (1997),
hep-th/9706109.}
\lref\vm{
J.~M.~Labastida and M.~Marino,
``Polynomial invariants for torus knots and topological strings,''
Commun.\ Math.\ Phys.\  {\bf 217}, 423 (2001), hep-th/0004196;\br
P.~Ramadevi and T.~Sarkar,
``On link invariants and topological string amplitudes,''
Nucl.\ Phys.\ B {\bf 600}, 487 (2001), hep-th/0009188;\br
J.~M.~Labastida, M.~Marino and C.~Vafa,
``Knots, links and branes at large N,''
JHEP {\bf 0011}, 007 (2000), hep-th/0010102;\br
J.~M.~Labastida and M.~Marino,
``A new point of view in the theory of knot and link invariants,''
math.qa/0104180;\br
M.~Marino and C.~Vafa,
``Framed knots at large N,''
hep-th/0108064.}
\lref\sheld{
S.~Kachru, S.~Katz, A.~E.~Lawrence and J.~McGreevy,
``Open string instantons and superpotentials,''
Phys.\ Rev.\ D {\bf 62}, 026001 (2000)
hep-th/9912151;\br
S.~Kachru, S.~Katz, A.~E.~Lawrence and J.~McGreevy,
``Mirror symmetry for open strings,''
Phys.\ Rev.\ D {\bf 62}, 126005 (2000)
hep-th/0006047.}
\lref\sheldii{S.~Katz and C.~M.~Liu,
``Enumerative Geometry of Stable Maps with Lagrangian Boundary Conditions and Multiple Covers of the Disc,''
math.ag/0103074.}
\lref\MayrHH{
P.~Mayr,
``On supersymmetry breaking in string theory and its realization in brane  worlds,''
Nucl.\ Phys.\ B {\bf 593}, 99 (2001),
hep-th/0003198.}
\lref\quintic{
I.~Brunner, M.~R.~Douglas, A.~E.~Lawrence and C.~R\"omelsberger,
``D-branes on the quintic,''
JHEP {\bf 0008}, 015 (2000)
hep-th/9906200.}
\lref\ilka{I.~Brunner and V.~Schomerus,
``On superpotentials for D-branes in Gepner models,''
JHEP {\bf 0010}, 016 (2000), hep-th/0008194.}

\vskip-2cm
\Title{\vbox{
\rightline{\vbox{\baselineskip12pt\hbox{CERN-TH/2001-230}
                            \hbox{hep-th/0108229}}}\vskip1.5cm}}
{N=1 Mirror Symmetry}
\vskip -0.99cm
\centerline{\titlefont and Open/Closed String Duality}

\abstractfont 
%
\vskip 0.8cm
\centerline{Peter Mayr}
\vskip 0.8cm
\centerline{CERN Theory Division} 
\centerline{CH-1211 Geneva 23}
\centerline{Switzerland}
\vskip 0.3cm
\abstract{%
We show that the exact $\cx N=1$ superpotential of
a class of 4d string compactifications is computed by
the closed topological string compactified to two dimensions.
A relation to the open topological string is 
used to define a special geometry for $\cx N=1$ mirror symmetry.
Flat coordinates, an $\cx N=1$ mirror map for chiral multiplets and 
the exact instanton corrected superpotential 
are obtained from the periods of a system of differential equations.
The result points to a
new class of open/closed string dualities which 
map individual string world-sheets with boundary to ones without.
It predicts an mathematically unexpected
coincidence of the closed string Gromov--Witten invariants of
one \CY geometry  with the open string invariants of the dual \CY.
}
\vskip1cm
\Date{\vbox{\hbox{ {August 2001}}
}}
\goodbreak

\parskip=4pt plus 15pt minus 1pt
\baselineskip=14pt 
\leftskip=8pt \rightskip=10pt
%

\newsec{Introduction}
It has been known that the exact non-perturbative holomorphic prepotential
that determines the effective action of an $\cx N=2$ effective
field theory, can be determined from the periods of a geometric 
object, both in the pure field theory \swp\ and in the low energy
string theory \netst\ context.  
The purpose of this note is to show that a similar statement
is true for the holomorphic superpotential 
\eqn\spi{
\int d^4x d^2\theta\ W(\Phi),
}
for a class of $\cx N=1$ effective string theories.
The argument will involve a new relation between
certain open and closed string backgrounds
which relates the superpotential \spi\ to an amplitude of the
closed string compactified on a $\hx c=4$
background to two dimensions.

The $\cx N=1$ supersymmetric situation that we consider 
arises in a type II \CY compactification with D-branes 
wrapped on cycles in the interior manifold and filling space-time. 
There is a closely related open topological string theory \wittopop\
which has been used in \vafaext\foot{See also \oo.} to 
propose an extension of mirror symmetry to include D-branes. 
A vital break-through in this direction has been made in the
paper by Aganagic and Vafa \AVi\ for a class of non-compact 
D-brane geometries. The exact instanton correct superpotential
has been obtained by a combination of closed string mirror
symmetry and open string methods \AVi; see also
\AVii\ for a further discussion and applications and
\quintic\phantom{\hori\sheld\ilka\sheldii}\hskip-50pt - \AViii\
for other related work in this direction.

We will follow a different route by observing that
the connection to the $\hx c=4$ topological closed string predicts
a certain geometry of the holomorphic F-terms inherited from the
2d amplitudes. This {\it $\cx N=1$ special geometry} leads to 
a definition of mirror symmetry for the chiral multiplets
which is completely analogous to $\cx N=2$ closed string
mirror symmetry. In particular we derive a system of
differential equations for the special geometry
that treats the chiral multiplets from the open and closed string 
sector on a complete equal footing. An
$\cx N=1$ mirror map for the chiral multiplets and the exact holomorphic 
$\cx N=1$ function $W(\Phi)$ are then defined by the 
solutions, or ``periods'', of this system, very similar to the
case of closed string mirror symmetry.

We will first recover the results of \AVi\ on the superpotential
from this
definition of $\cx N=1$ mirror symmetry. To appreciate that this is
not merely a technicality, note firstly that by treating 
the closed and open string moduli on the same footing, we
obtain a complete, global description of the holomorphic $\cx N=1$ moduli
space\foot{That is moduli modulo D-terms.}, 
including phase transitions, singularities and
the associated analytic continuations and monodromies. 
Also this approach to $\cx N=1$ mirror symmetry appears
to be a framework for generalizations
to many D-branes and higher genera amplitudes.

A central argument will be the new relation
between open and closed string backgrounds which identifies
certain topological amplitudes of the open string compactification with central
charge $\hx c=3$ and a ``dual'' closed string with $\hx c=4$. 
These topological 
amplitudes are related to physical space-time amplitudes in the type II 
string on a \CY 3-fold with D-branes and a type II string on a \CY 4-fold
without branes but with certain fluxes, respectively.
We will separate the case of disc and sphere world-sheet topologies,
where a relation of the above kind will be verified, from the case 
of general world-sheet topologies which will remain at the level 
of a speculation. It will be interesting to study the conjectural 
open/closed string duality in other world-sheet topologies,
both from the mathematics and physics point of view.
If true, an infinite number of holomorphic four-dimensional $\cx N=1$ 
amplitudes would be computed by the 2d string theory.
Some evidence in favor of a true open/closed string duality 
will be given by studying M-theory on
a nine-dimensional manifold.

A mathematically quite surprising connection implied 
by the genus zero\foot{More precisely open $g=0,h=1$ vs. closed $g=0$.}
duality  is one 
between the moduli spaces of discs in 
the D-brane geometry on a \CY 3-fold $Y$, and the moduli spaces 
of spheres in the dual \CY 4-fold $X$.
Specifically the closed string Gromov--Witten invariants of $X$, 
which count the appropriately defined number of holomorphic 
spheres \tfii, and the 
integral open string invariants \ov\ of the D-brane geometry $(L,Y)$, 
which count the number of discs in $Y$ with boundary on the D-brane, coincide
for a dual pair!
In other words the open/closed string relation maps {\it individual}
open string world-sheets with boundary to those without\foot{This 
is similar to a duality described in \AViii.}.

In this paper, we will discuss the basic ideas and arguments
outlined above, mostly for the type IIB D-branes.
Other aspects, such as the global structure of the
$\cx N=1$ moduli space, phase transitions, the type IIA (D-brane) geometry, 
more explicit calculations and various generalizations 
will be discussed in \wip. 

The organization of this paper is as follows. In sect. 2 we formulate
the precise proposal for the relation between the topological 
amplitudes at genus zero of two string backgrounds with and without branes.
We explain why the $\hx c=4$ closed topological string computes the
4d $\cx N=1$ superpotential \spi. In sect. 3 we identify pairs
of closed and open string backgrounds which are supposedly related
by a duality. We verify their equivalence at the level of the 
genus zero amplitudes related to the superpotential (a calculation
of period integrals is relegated to app. A for readability.).
In sect. 4 we use the established ($g=0$) connection between the 
open and closed topological string theories to define $\cx N=1$
mirror symmetry for scalars of chiral multiplets in a way very
similar to closed string, $\cx N=2$ mirror symmetry. The central
element is a system of differential equations, whose solutions,
or periods, define the flat coordinates, the $\cx N=1$ mirror map
and the 4d superpotential. The approach is illustrated with a 
simple example, including an explicit demonstration of the 
agreement of the closed string (sphere) and open string (disc)
Gromov--Witten invariants in the dual  backgrounds. In sect. 5
we provide some evidence for an extension of the open/closed string
relation to a true string duality, by studying the M-theory limit
of the type IIA compactifications.

\newsec{Topological string amplitudes in type II \CY compactifications}
In this section we review the space-time interpretation 
of various topological string theories on a \CY background 
and formulate the precise proposal for a coincidence of 
certain genus zero topological amplitudes of a open string
theory on a \CY 3-fold and a conjecturally dual closed string
on a \CY of one dimension higher.

\subsubsec{Topological closed $\cx N=2$ amplitudes and 3-cycle integrals}
The $\cx N=2$ closed topological string \wittop\ on a \CY 3-fold 
has non-zero partition function $\cx F_g$ for all world-sheet genera $g$.
At $g=0$ the topological string computes the holomorphic prepotential
of the four-dimensional effective $\cx N=2$ string theory. For B-twist,
the result may be expressed by the geometric period integrals \spgeo\ 
of the holomorphic 3-form over special Lagrangian 3-cycles in the \CY $Y^*$
\eqn\pintii{
\Pi(\cmm z_a)=\int_{C_3} \OM3(Y^*),\qquad C_3 \in H_3(Y^*),}
from which $\cx F_0$ can be recovered by integration.
As indicated the period $\Pi$ depends on the 
complex structure moduli $\cmm z_a$ of $Y^*$ only. 
The superscript will be used to indicate that these moduli 
arise in the closed string sector.
Although the expression \pintii\ looks completely classical,
it comprises a highly intricate sum of instanton corrections. This is best seen
in the equivalent A-twisted model on the mirror manifold $Y$, where
these instantons arise from Euclidean string 
world-sheets wrapped on holomorphic
2-spheres. Most of the complexity of the geometric expression \pintii\ 
is hidden in the relation between the vev's of 
physical scalar components $\cmm t_a$ and the geometric moduli $\cmm z_a$,
which is given by a ratio of periods
\eqn\csmm{
\cmm t_a(\cmm z_b)=\Pi_a(\cmm z_b)/\Pi_0(\cmm z_b),
}
and is called the mirror map. Geometrically, the real scalars
$\rmx{Im}\, \cmm t_a$ measure the size of holomorphic spheres in $Y$ on which 
the type IIA world-sheet instantons wrap; as the BPS action of 
these instantons is proportional to the area of the sphere, 
the instanton weight is $\sim q_a=e^{2\pi i \cmm t_a}$. Knowing the mirror map
\csmm, one obtains from the remaining periods $\tx \Pi_a$
the first derivatives of the prepotential $\cx F_0$, the latter having an
integral instanton expansion in the large $t$ limit of the form \tfii:
\eqn\wclexp{
\tx \Pi^{inst}_a=\p_a\cx F_0=\sum_{\vx n} n_a D_{\vx n}\sum_{m=1}^\infty 
\fc{\vx q^{m\vx n}}{m^2}.
}
Here $q^{\vx n}=\prod_a q_a^{n_a}$ and the integral coefficients $D_{\vx n}$
are the Gromov--Witten invariants which count the appropriately 
defined number of holomorphic spheres in $Y$.

\subsubsec{Topological open $\cx N=1$ amplitudes and 3-chain integrals}
One may add D-branes to the previous compactification on the
\CY 3-fold such that supersymmetry is broken to $\cx N=1$.
This situation is described by the $\cx N=1$ open topological string theory 
\wittopop. The partition function  
has an expansion $\cx F=\cx F_{g,h}$ in world-sheets of genus $g$ and 
with $h$ holes. The holomorphic functions $\cx F_{g,h}$ of the A-twisted
model are related to $d=4$ $\cx N=1$ superpotential
terms on the world-volume of a D6-brane wrapped on a special Lagrangian
3-cycle $L$ in $Y$ \ksp\ov\vafaln:
\eqn\fdtosa{
h\, \int d^4x d^2\theta \cx F_{g,h} (\cx W^2)^g\, (\cx F^2)^{h-1},
}
where $\cx W$ is the gravitational chiral superfield 
and $\cx F$ is the 
chiral superfield for the gauge supermultiplet on the D-brane. 
In particular the partition function $\cx F_{0,1}$ 
computes the superpotential \spi. 
There is a formula similar to \pintii\ that describes the same superpotential
in terms of the mirror D-brane geometry which is, 
for appropriate choice of $L$, a D5-brane wrapped on
a 2-cycle $C$ in the mirror manifold $Y^*$ \AVi\foot{See also 
\Witbranes\sheld.}
\eqn\pinti{
W = \int_{D_3} \Omega = \int_{\p D} \omega = W(\cmm z_a,\omm z_i).
}
Here $D_3$ is a 3-{\it chain} in $Y^*$ with $C$ as a boundary component 
and $\Omega$ is the holomorphic (3,0) 
form on $Y^*$, which may locally be written as $\Omega=d\omega$.
Due to the B-twist,
$W(C)$ depends only on the complex structure moduli $\cmm z_a$ 
of $Y$ and the moduli $\omm z_i$ of the 2-cycle $C$. 
Note that the former are from the closed string sector 
while the latter are moduli of the open string sector.

Again, the geometric expression $W$ describes highly delicate
instanton physics which is best
seen in terms of {\it disc} world-sheet instantons in the type IIA
geometry $(Y,L)$. Knowing the relation between the 
scalar vev's $(\cmm t_a;\omm t_i)$ of the chiral multiplets and the 
geometric moduli $(\cmm z_a;\omm z_i)$, the instanton 
expansion may be extracted from the geometric integral \pinti,
which is predicted \ov\ to have an integral large $t$ expansion
of the form:
\eqn\wopexp{
W=\sum_{\vx k,\, \vx m }
d_{\vx k,\vx m}\, \sum_n \fc{\vx 
q_{(c)}\, ^{n \vx k}\, \vx q_{(o)}\, ^{n \vx m}}{n^2},
}
where $q_{(c)}=e^{2\pi i \cmm t_a}$ ($q_{(o)}=e^{2\pi i \omm t_a}$) 
are the exponentials of 
the (complexified) volumes of holomorphic spheres (holomorphic discs) in a CY
3-fold $Y$. Moreover, the expansion coefficients 
$d_{\vx k,\vx m}$ are the integral open string invariants 
that count the appropriately defined 
number of discs in the class $(\vx k,\vx m)$.
The integrality predictions of \ov\ have been 
verified by now in a large number of highly non-trivial 
examples \AVi\AVii\sheldii\AViii\vm.

\subsubsec{Topological closed ``$\cx N=1$'' amplitudes}
A topological closed string compactification with $\hx c=4$ maybe 
defined by a \CY 4-fold $X$. This is related to a 2d  space-time
theory with the same number of supercharges as $\cx N=1$ in $d=4$.
The two different topological field theories can be used to study mirror
symmetry for 4-folds \gmp\ffsp, with the details being slightly
different from the ``critical'' case of \CY 3-folds due to different
ghost number of the vacuum. The basic genus zero amplitude
is the topological triple coupling 
$C_{\al\be\ga}=\langle \cx O^{(1)}_\al\cx O^{(1)}_\be \cx O^{(2)}_\ga\rangle$,
where $\cx O^{(i)}$ is an operator localizing on a codimension $i$
hypersurface in $X$. They are related to the periods of the
holomorphic 4-form of the mirror $X^*$
\eqn\ffper{
\Pi_\al(\cmm z_\be)= \int_{\ga_\al}\OM4,\qquad  \ga_\al\in H^{\Omega}_4(X^*),
}
where the superscript $\Omega$ denotes the 
sub-sector of homology 4-cycles which has a representative 
calibrated by the holomorphic $(4,0)$ form.
Specifically the topological triple couplings $C_{\al\be\ga}$ are the 
double derivatives \ffsp
\eqn\ffp{{
C_{\al\be\ga}\ =\ \fc{\p}{\p t_\al}\fc{\p}{\p t_\be}\, \Pi^{\ln^2}_\ga}} 
of the middle periods with leading double logarithmic behavior in the 
large $t$ limit. Moreover these periods have the 
integral large $t$ expansion \gmp\ffsp
\eqn\ffpi{
\Pi_\ga^{\ln^2} \ =\ P^2_\ga(t_\al)+
\sum_{\vx k} D^\ga_{\vx k}\, \sum_n \fc{\vx q\, ^{n \vx k}}{n^2},}
where $P^2_\ga$ is a certain degree two polynomial. Note that this expansion is
precisely of the same form as the $\cx N=1$ open string expansion
\wopexp. 

\subsubsec{$\underline{\rmx{The\  proposal}}$}

\vskip-14pt
Our proposal is that for appropriate choice of the open string
D-brane geometry $(Y,L)$ and a dual closed string geometry 
$X$, the open and closed string expansions \wopexp\ and 
\ffpi, are identical. This is the same as saying that the
topological amplitudes of the open string at $g=0,h=1$ defined on 
the D-brane geometry $(Y,L)$ and the closed string at $g=0$ on $X$ 
are identical.

We will identify appropriate duals  $(Y,L)$ and $X$ in the next 
section and verify the agreement of the genus zero topological 
amplitudes. Clearly it will be extremely interesting to see 
whether a similar relation exists for other world-sheet topologies
and the relation at genus zero extends to a true 
string theory duality \wip. An M-theory argument
in favor of this conjectural string duality will be
given in sect. 5\yyy.

The coincidence of the $g=0$ $(h=1)$ open/closed string amplitudes 
has the following {\it two-dimensional} space-time interpretation. 
Consider the D-brane geometry $(Y,L)$, however with
a D4-brane wrapped on the 3-cycle 
$L$ instead of a D6-brane. The open topological
amplitudes $\cx F_{g,h}$ compute the following terms in the
2d effective world-volume \ov:
\eqn\ovii{
h\, \int d^4x d^2\theta\delta^2(x)F_{g,h}(\cmm t_a;\omm t_i)
(\cx W^2)^g(\cx W\cdot v)^{h-1},}
where the delta function localizes on the D4 world-volume and the
tensor $v$ points into the orthogonal directions. For $g=0,h=1$
this is the two-dimensional superpotential in a theory with 4 supercharges.
Precisely the same superpotential arises in the type II \CY
4-fold compactification with background 
fluxes \gvw\foot{See also \wolfg\jan.}:
\eqn\fff{
W=\sum N_a \int_{\gamma_a} \OM4,\hskip30pt \gamma_a \in H_4(X^*).}
Here $N_a$ are integers that specify the background fluxes.
Moreover we have used the type IIB language on the mirror $X^*$ 
for simplicity; see 
\gvw\vafaln\ for the corresponding type IIA superpotential
on $X$.
The superpotentials \ovii\ and \fff\ are defined in the same
class of 2d theories with four supercharges. The conjecture says that 
for the appropriate choice of $Y$, $L$ and $X$, they are
identical. Note that the flux on the 4-fold is important
for the duality to work. This is very similar as for
the AdS/CFT correspondence \mal\ and Vafa's large $N$ duality \vafaln%
\foot{In fact background fluxes in type II compactifications
in \CY {\it 3-folds} give also rise to superpotentials in
four-dimensional compactifications with a perturbative $\cx N=1$ 
supersymmetry \TV\ssb. This is the only known other case where 
an infinite sum of instanton corrections to a $\cx N=1$
superpotential can be computed systematically, by conventional
$\cx N=2$  mirror symmetry \TV\ssb\vafaln\CIV.}%
.

To be more precise, as we consider non-compact manifolds, 
some of the periods \ffper\ are defined on non-compact
cycles. The moduli dependent contribution of these periods to the 
superpotential \fff\ is described by the {\it variation} of the volume of 
the non-compact cycle with the complex structure of 
$X^*$. Rather than in terms of flux on the non-compact cycle,
this contribution is better thought of as the response 
to variations of the complex structure relative to a fixed behavior at 
infinity, similarly as in \AVi\Witbranes.

Note that, once the closed topological amplitude \ffpi\ is 
identified with $\cx F_{0,1}$ by the D4-brane interpretation \ovii,
it translates immediately 
to the holomorphic superpotential $W$ in the 
{\it four-dimensional} $\cx N=1$ theory, by the alternative
space-time interpretation \fdtosa\ of $F_{0,1}$ on the D6-brane world-volume. 
It is by this chain of 
arguments that the 2d closed string 
may be seen to compute 4d holomorphic $\cx N=1$ space-time
couplings.

It is worth stressing that the integral coefficients 
of the identified expansions, $d_{\vx k,\vx m}$ in
\wopexp\ and 
$D_{\vx k}$ in \ffpi, have vastly different meanings.
The first counts the appropriately defined number of discs in 
the D-brane geometry $(Y,L)$, while the other 
the appropriately defined number spheres in $X$. This is the
aforementioned new feature of this class of  open/closed string ``dualities'':
world-sheets with boundaries are mapped individually to
world-sheets without.

Similarly, the duality identifies the closed and open string moduli 
$(\cmm t_a;\omm t_i)$ of the D-brane geometry $(Y,L)$ with
only the closed string moduli
$\cmm t_\al$ of $X$. The first are defined as the areas of 
spheres and discs in $Y$, respectively, while the latter 
measure only volumes of spheres in $X$.

\newsec{Open/closed string duals in two dimensions}
While it has been argued in the previous section that the
open topological amplitude $\cx F_{0,1}$ in the D-brane geometry
$(Y,L)$ {\it can} in principle coincide with the topological genus zero
amplitude on $X$, it is the purpose of this section to identify
the appropriate $X$ for a given type IIA D6-brane configuration $(Y,L)$.
In fact it is more convenient to use the equivalent mirror geometries and
to identify a \CY 4-fold $X^*$ for a given type IIB D5-brane 
geometry $(Y^*,C)$\foot{The type IIA mirror geometries will be
discussed in \wip.}.

Recall that the
D-brane geometry consists of a D5-brane wrapped on the 2-cycle
$C$  of a \CY 3-fold $Y^*$, with the closed string moduli for the
\CY geometry related to the 3-cycle integrals \pintii\ and the 
open string moduli for the geometry of the D-brane related to
the 3-chain integrals \pinti.
An appropriate formulation of the problem is to find for
a given geometry $(Y^*,C)$ a 4-fold $X^*$ such that there is 
an injective and surjective map 
\eqn\georel{
H_3(Y^*,C)\to H^{\Omega}_4(X^*),
}
where $H_3(Y^*,C)$ denotes the relative 
homology modulo boundaries on $C$.
This asserts that all 3-cycle \pintii\ and
3-chain \pinti\ integrals in $Y^*$ may be associated with 
an appropriate 4-cycle integral \ffper\ in $X^*$.\foot{As alluded
to earlier, the manifold $X^*$ and some of the 4-cycles will be
non-compact.}

\subsubsec{The triple $(Y^*,L,X^*)$ in the 2d linear sigma model}
In the following we will use the 2d linear sigma model \wlsm\ to
study the relevant \CY geometries with and without D-branes.
The appropriate 4-fold $X^*$ for the closed string compactification
may be constructed from the linear sigma model for the 
3-fold $Y^*$ by a simple procedure that adds a 
{\it linear 2d superpotential in a new variable $v$}.
Concretely we consider non-compact \CY $d$-folds defined by a 
2d superpotential\foot{We use $\Wtd$ to denote the 2d {\it world-sheet}
superpotential of the linear sigma model. This should not be confused with
the space-time superpotential $W$ in two dimensions.}
\eqn\mirrorsp{
\Wtd=\sum_{i=0}^{d-1+h_{1,d-1}} a_i y_i,
\qquad \prod_i y_i^{\lm a_i}=1,\qquad a=1,...,h_{1,d-1},}
where the variables $y_i$ take value in $\CC^*$ and the $a_i$ are
constants that parametrize the complex structure. 
Moreover the $\lm a$ are $h_{1,d-1}$ linearly independent
vectors with integral entries that 
define the specific \CY manifold\foot{The vectors $\lm a$ are
also the
charge vectors of the type IIA gauged linear sigma model on the mirror manifold.}.
For further
background on the definition of \CY manifolds and mirror symmetry
we refer to \HV\HIV.

By rescalings of the $y_i$, the superpotential $\Wtd$
depends only on $h_{1,d-1}$ combinations of the $a_i$,
which is the dimension of the complex structure moduli 
space $\cx M_{CS}$.
A canonical choice of coordinates on this space 
is provided by the so-called algebraic 
coordinates $\cmm z_a$ defined as 
\eqn\defz{
\cmm z_a=\prod_i a^{\lm a_i}_i.}
There will in general be several proper 
choices of linear combinations of the $\lm a$ which define 
good coordinates on various patches of $\cx M_{CS}$. 

After solving for the relations in \mirrorsp, the 
superpotential depends on $d$ variables
for a \CY $d$-fold, $\Wtd=\Wtd(y_0,...y_{d-1})$. An equivalent
definition \HIV\ is in terms of the hypersurface\foot{See app. 
A\yyy for the derivation.}
\eqn\tdspii{\Wtd(\ty_0,...,\ty_{d-2})+xz=0,\qquad \ty_i=\fc{y_i}{y_{d-1}}.}

We are interested in a D5 brane on a 2-cycle $C$ in the 3-fold $Y^*$,
such that there is an $\cx N=1$  world-volume theory on it with 
classically zero, but non-perturbatively (in the world-volume sense) non-zero 
superpotential $W$. For holomorphic $C$ , the superpotential \pinti\ is 
identically zero. A non-zero brane superpotential arises 
for a non-compact 2-cycle with fixed behavior at infinity \Witbranes\AVi.
An appropriate $C$ in the hypersurface \tdspii\ is defined by
\eqn\defC{
C:\  x=0, \ \ty_i=\ty_i(r),}
where $z$ is the holomorphic coordinate on $C$, $r=|z|$
and near infinity, the coordinates $\ty_i,\, i=0,1$ approach a fixed value
$\ty_i^\infty$. The 4d superpotential $W$ depends only on the
values of the $\ty_i$ at $r=0$ which may be specified by the
ratio 
\eqn\defzn{
\omm z_0=\ty_0/\ty_1|_{r=0}.}
Eventually the superpotential $W$ for the D5 brane on 
$C$ evaluates to \AVi
\eqn\spavi{
W(\cmm z_a;\omm z_0)=
\int_{\ty_1^\infty}^{\ty_1|_{r=0}} \ln(\tx y_0) d\ln(\tx y_1),
}
where the value of $\ty_0$ is fixed in terms of $\ty_1$ by the fact that
$\Wtd$ vanishes on $C$.
Note that $\omm z_0$ is the 
relevant open string modulus for the D-brane geometry. 
It is related to the scalar component of a chiral multiplet by a
field redefinition described below.

We may finally formulate the \CY 4-fold $X^*$ on which the conjecturally 
dual closed string is compactified.
Let $\Wtd(Y^*)$ denote the 2d superpotential for the \CY 3-fold 
$Y^*$ on which the 
previously described D-brane geometry is defined. The 2d linear sigma model
for  $X^*$ is defined by the superpotential and relation
\eqn\defX{
\Wtd(X^*)=\Wtd(Y^*)+ a_1 v_1+ a_2 v_2,\qquad
y_0v_1=y_1v_2.}
Here $v_i$ are two new variables in $\CC^*$. 
The $h_{13}(X^*)=h_{12}(Y^*)+1$ charge vectors $\lm a$ for $X^*$ 
are 
\eqn\defzt{\eqalign{
\lm {0}&=(1,-1,0,\dots;1,-1),\cr
\lm {a}&=(\hskip20pt\lm a_Y\hskip15pt;\, 0\, ,\ 0),\hskip 55pt a=1,...,h_{13}-1,
}}
where the last two entries refer to the new fields $v_i$.
The $\lm a$ define $h_{13}$ coordinates \defz\ on the
complex structure moduli space $\cx M_{CS}(X^*)$. 
In the duality to the D-brane geometry $(C,Y^*)$, the complex 
structure moduli of $X^*$ will be identified with the moduli
of the D-brane geometry $(C,Y^*)$ as described in Tab. 1\yyy: 

\vbox{
\vskip 0pt
\eqn\modids{
\vbox{\offinterlineskip\tabskip=0pt\halign{
\strut#&$#$~\hfil&\quad\vrule#\quad&$#$~\hfil&#\cr
\hskip40pt$Y^*\ (4d)$&&&&\hskip15pt $X^*\ (2d)$ \cr
\noalign{\hrule}
&&&&\cr
complex structure\ & \cmm z_a&&z_a, {\scriptstyle a>0} 
&\ \ complex structure\cr
D5-brane geometry\ & \omm z_0&&z_0& \hskip 31pt `` \cr}}}}
\noindent{\ninepoint  
{\bf Tab. 1\yyy}: The identifications of the parameters in the effective
4d $\cx N=1$ compactification with the parameters of the related 2d 
closed string compactification.}

\vskip 13pt\ni
It is instructive to have an intuitive picture of the
geometry of $X^*$. If we add
a further term $\sim v^{-1}$ to the superpotential $\defX$, 
then $X^*$ is a fibration of $Y^*$ over a cylinder 
$\CC^*_v$ parameterized 
by the variable $v$. There are also branch points on $\CC^*_v$ where the
fiber degenerates. This is sketched in Fig. 1\yyy.
In the limit where the coefficient of the term
$v^{-1}$ goes to zero, and we recover $X^*$, the cylinder becomes
infinitely long.
{\goodbreak\midinsert
\centerline{\epsfxsize 3truein \epsfbox{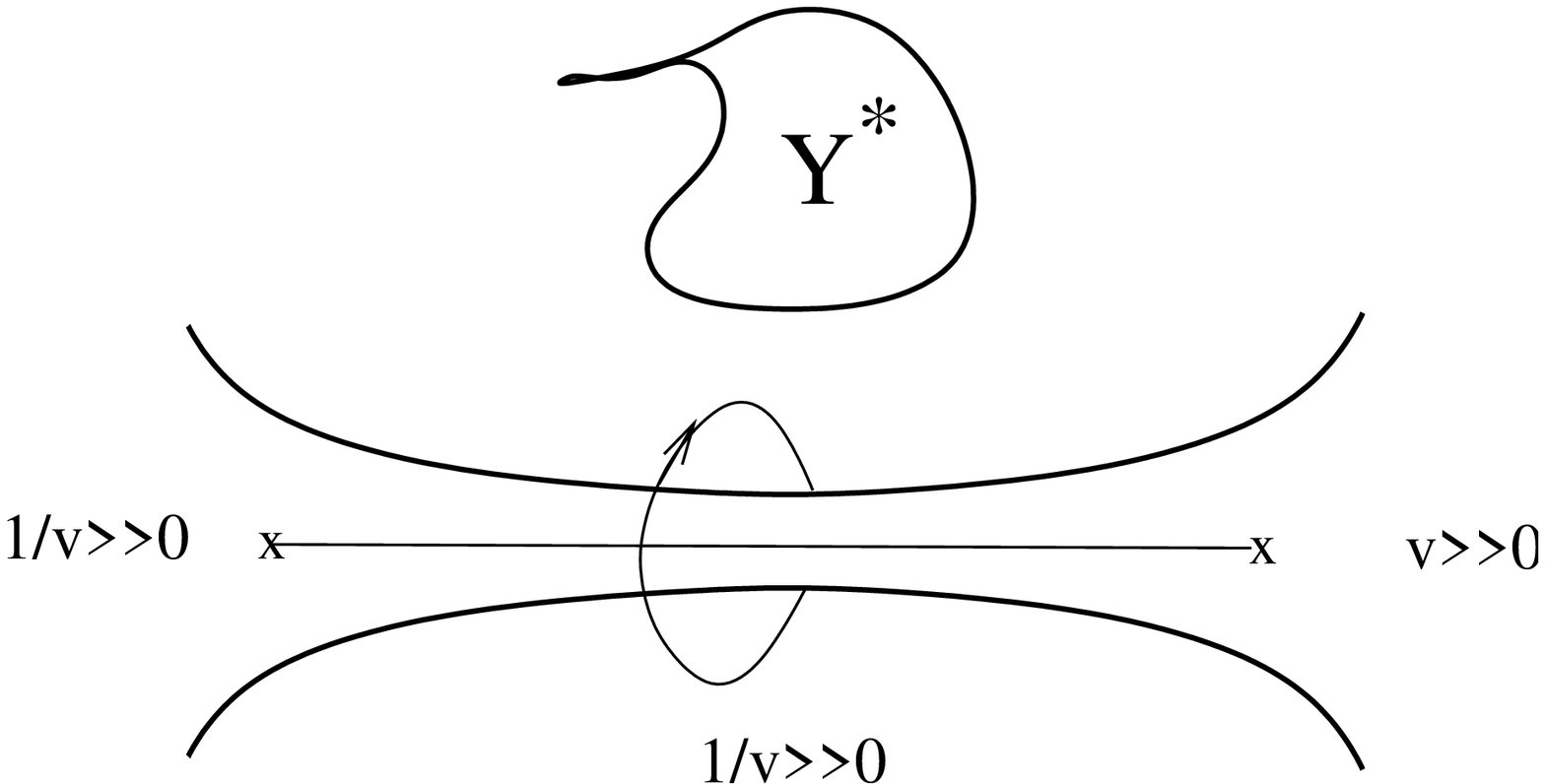}}
\leftskip 1pc\rightskip 1pc 
\noindent{\ninepoint  
{\bf Fig. 1\yyy}: The Calabi--Yau manifold $X^*$ is a ``fibration'' of 
$Y^*$ over a cylinder with branch points, in the limit where the
cylinder becomes infinitely long. 
}\endinsert}
There are two types of 4-cycles\foot{See also \netst\ for a very
similar discussion and \kbmv\ for closely related \CY geometries
and a connection to the Jacobi map.}
in $X^*$, made from 3-cycles in $Y^*$
and a closed cycle on $\CC^*_v$.
The ones of the first type are of topology $S^1\times C_3^{Y^*}$ 
and arise from pulling a 3-cycle in $Y^*$ around the periodic direction 
of the cylinder. The period integrals of the holomorphic 4-form 
evaluated on such a cycle reduce to the periods of the holomorphic
3-form in $Y^*$, times an irrelevant constant. 
The other type of cycles is obtained by integrating between branch
points on $\CC^*_v$ where a 3-cycle in the fiber vanishes. The
topology of such a cycle is in general $S^4$. 
We show in app. A that the period integral over this type of 4-cycle 
that is well-defined in the limit of the very long cylinder,
agrees with the {\it 3-chain} integral \spavi\ in $Y^*$. This
establishes the relation \georel\ and justifies the identifications 
made in \hbox{Tab. 1\yyy}.

\newsec{Special geometry for $\cx N=1$ mirror symmetry}
The 2d topological field theory amplitudes define a 
certain special geometry with a flat connection for the 
bundle of supersymmetric ground-states 
in special coordinates \ksp\gmp. This is a 
generalization of the well-known case of $\cx N=2$ special geometry
which follows from the so-called $tt^*$ equations \ksp.

Although there is no known physical duality in four dimensions,
it follows from the arguments in sect. 2\yyy that
the $\cx N=1$ 4d F-term inherits this structure from the 2d amplitudes.
This ``$\cx N=1$ special geometry'' leads to a definition of 
4d $\cx N=1$ open string mirror symmetry
which is completely analogous to that of $\cx N=2$ closed string
mirror symmetry. In the following derive a system
of differential equations for the $\cx N=1$ special geometry
and use it to describe 4d $\cx N=1$ mirror symmetry
of the chiral multiplets in terms of the solutions, or ``periods''
of this system. 

Note that the same differential equations describe
also the true periods of the two-dimensional type IIA compactified
on the 4-fold $X$, although this theory is not physically equivalent
(as it is defined in 2d).
The system of differential equations satisfied by the 
periods $\Pi(X^*)$ of the \CY 4-fold geometry for the 2d compactification
is of a generalized hyper-geometric, so-called GKZ type 
\cyper\foot{See also \hkty\ for a nice discussion of
the similar GKZ systems of \CY 3-folds.}.
This has been discussed in the study of mirror symmetry for 4-folds \gmp\ffsp.

It is important to note that the only data on which the special geometry
depends, are the charge vectors $\lm a$ in \defzt. 
Once the relation between the D-brane geometry 
$(C,Y^*)$ and $X^*$ described in the last section is understood,
one may immediately define the GKZ system for $\cx N=1$ mirror 
symmetry from the classical D-brane geometry  $(C,Y^*)$ 
in the following way. Let $C$ be
the D5-brane on $Y$ with the asymptotic behavior
\eqn\defc{
C: x=0,\qquad y_i=z_0\, y_j\ {\rm at}\ r=0,
}
such that $z_0$ is small and $z_0\to 0$ describes the
classical limit on the D-brane world-volume.
This is simply the appropriate definition of the D-brane in
the coordinates adapted to the relevant
patch in $Y^*$, where the D-brane is located. The D5-brane 
with the above asymptotic behavior defines the following
basis of charge vectors for $\cx N=1$ mirror symmetry:

\vbox{
\vskip 10pt
\eqn\modids{
\vbox{\offinterlineskip\tabskip=0pt\halign{
\strut
\hfil$#$~
&$#$~\hfil
&\hskip 5pt$#$\hskip 5pt
&\hskip5pt$#$\hskip10pt
&\hfil$#$~&\hfil$#$~&\hfil$#$~&\ \ \ $#$\cr
\lm {a}&=(&&\lm a_{Y^*}&;&0&0\, ),&\ \ a=1,...,h_{12}\cr\cr
\lm {0}&=(0,&...,0,1,0,...&...,0,-1,0,...&...,0\, ;&1&-1\, ).&\cr
&&{\scriptstyle i-th\ position}&\ {\scriptstyle j-th\ position}\cr
}}}}

\ni
The rank of the set of charge vectors of the above form equals
the rank of the lattice of all possible supersymmetric D5-brane 
charges on $Y^*$. In other words any D5-brane which is mirror 
to a D6-brane on a special Lagrangian 3-cycle of topology 
$\CC\times S^1$ may be described by the charge vectors \modids\
for appropriate $i$ and $j$. The good local coordinates \defz\ are 
then defined by a linear combination of the $\lm \al$ within a given basis 
\modids. The precise linear combination is obtained by finding the
minimal generators of the K\"ahler cone as will be explained
in more detail in \wip; see also the example below for a brief explanation.

For ease of notation we will then use the form $\defzt$ corresponding to 
$i=0,j=1$ in the following; the general case is recovered by 
a trivial redefinition of coordinates. 
The searched for differential equations may be 
straightforwardly obtained from the LG definition of
the periods on $X$ \cec 
\eqn\lgpers{
\Pi(X^*)\sim\int \prod_0^{d-1} \fc{dy_i}{y_i}\ e^{-\Wtd(X^*)}.}
From \mirrorsp\ it is easy to see that they
satisfy the differential equations $\cx D_{\al}\, \Pi=0$ with 
\eqn\defGKZ{
\cx D_{\al}=\prod_{\lm \al _i>0}\big(\fc{\p}{\p a_i}\big)^{\lm \al _i}-
\prod_{\lm \al _i<0}\big(\fc{\p}{\p a_i}\big)^{-\lm \al _i},
}
where $\lm \al,\, \al=0,\dots,h_{12},$ are the charge vectors. 
The same operators
rewritten in terms of logarithmic derivatives 
$\theta_\al=z_\al\fc{\p}{\p z_\al}$ of the proper coordinates \defz\ are
\eqn\defGKZii{
\cx D_\al=
\prod_{\lm \al_i>0}\prod_{j=0}^{\lm\al_i-1}
\big(\sum_\be \lm\be_i\theta_\be-j \, \big)-
z_\al\, 
\prod_{\lm\al_i<0}\prod_{j=0}^{-\lm\al_i-1}
\big(\sum_\be \lm\be_i\theta_\be-j\, \big).}
From the charge vector $\lm 0$ in \defzt\ we obtain the following
operator $\cx D_0$ for the open string modulus $\omm z_0$:
\eqn\defDn{
\cx D_0 = (1-\zo)\theta_0^2\, +\, 
\sum_{b>0}\, (\lm0_b+\zo\, \lm1_b) \theta_b\theta_0.}
In particular $\cx D_0$ consists of only terms of second degree.

The classical limit, which defines the instanton expansion, 
corresponds to small $z_\al$; in particular the leading 
behavior of the relation between the physical scalar fields
$t_\al$ in the chiral multiplets 
and the geometric parameters $z_\al$ in the 2d world-sheet superpotential
is $t_\al \sim \fc{1}{2\pi i}\ln(z_\al)$. The classical 
limit corresponds to a so-called point of maximally unipotent
monodromy\foot{See e.g. \agm.}, 
which also implies that the leading behavior of the
solutions of the GKZ
system expanded around the classical point $z_\al=0$ are in agreement
with the physical expections. Concretely there is one holomorphic solution,
which is just a constant 
$$
\Pi^{(0)}=1,
$$
as follows from the fact that the $\cx D_\al$ do not have a constant piece.
This is a consequence of the non-compactness of $X^*$.

The next set of solutions have a single logarithmic leading behavior
\eqn\sls{
\Pi^{1}_\al=\fc{1}{2\pi i}
\ln(z_\al)+S_\al(z_\be), \qquad \al,\be=0,...,h_{12},
}
where $S_\al$ are a power series in the geometric parameters $z_\be$.
These periods specify the instanton corrections to the classical relations 
$t_\al \sim \fc{1}{2\pi i}\ln(z_\al)$;
indeed  $z_\al=e^{2\pi i t_\al}+...$ is the weight of the string world-sheet
instanton wrapped on a holomorphic 2-cycle with area $\rmx{Im}\, t_a$.
The above periods define the flat coordinates and the 
\eqn\neomm{
\underline{N=1\ mirror\ map:}\hskip 40pt
t_\al(\omm z_0;\cmm z_a)=\fc{\Pi_\al^{1}(\omm z_0;\cmm z_a)}
{\Pi^{0}(\omm z_0;\cmm z_a),}}

\ni
where $\al$ runs over the indices of both the 
open and closed string moduli. The mirror map \neomm\
describes the exact functional dependence of the scalars vev's $t_\al$ of
the chiral $\cx N=1$ multiplets on the geometric open
and closed string moduli $z_\al=(\omm z_0;\cmm z_a)$.

The $\cx N=1$ mirror map \neomm\ has a peculiar property inherited
from the special set of charge  vectors \modids.
It has already been mentioned that there is no constant term in all 
of the operators $\cx D_\al$. The logarithms in the ansatz \sls\
produce simple source terms in the differential equation for 
the corrections $S_\al$:
$$
\cx D_\al S_\be + z_\al\, A^\al_\be = 0,
$$
where the $A^\al_\be$ are integers that characterize the linear part 
$\cx D_\al^{lin}=z_\al\, \sum A^\al_\be\theta_\be$ of the operators 
$\cx D_\al$.

From the special form of the operator $\cx D_0$ in \defDn\ 
it follows immediately that the 
source terms $A^0_\be$ are zero and the power series
$S_\al(z_\be)$ are independent of the open string modulus $\zo$.
This is in full agreement with the result of \AVii\ derived from 
open string methods. Moreover, it follows from the above
that the correction $S_0(z_b)$ is given by the following linear 
combination of the $S_a(z_b),\ a,b>0$:
$$
S_0 = \sum_{b=1}^{h_{13}-1}r_b S_b,\qquad  
r_b=(A_b^a)^{-1}\, A^a_0.
$$
Here we have assumed that the matrix $A_a^b,\ a,b>0$
is invertible. In the degenerate case there are further relations 
between the corrections $S_a$ with $a>0$ and a similar reasoning applies after
choosing a linearly independent basis of power series $S_a$.

The next class of solutions has a double logarithmic leading behavior
\eqn\dlper{
\Pi^2_\al=c_\al^{\be\ga} t_\be t_\ga +b_\al^\be t_\be+a_\al
+\cx O(e^{2\pi i t_\be})
}
where the coefficients $a_\al,b_\al,c_\al$ are constants and we have inverted
the mirror map \neomm\ to eliminate the $z_\al$ in favor of the 
physical fields $t_\al$. These periods encode the topological 
triple couplings \ffp\ and the Gromov--Witten invariants \ffsp.
In order that the ansatz \dlper\ solves
the operator $\cx D_0$, the leading coefficients $c_\al$ must obey
$$
c_\al^{00}+\sum \lm a_0c^{0a}_\al=0.
$$
The solution of this equation with non-zero $c_\al^{00}$ determines the 
period $\Pi^2_W$ that describes the\br\br
$\underline{ N=1\ superpotential:}$
\eqn\neosp{W^{\cx N=1}_{d=4}(\omm t_0;\cmm t_a)=
W^{\hx c=4}_{d=2}(\omm t_0;\cmm t_a)=
\fc{\Pi_W^{2}(\omm t_0;\cmm t_a)}
{\Pi^{0}(\omm t_0;\cmm t_a)}.}

\subsubsec{An example: $\cx N=1$ mirror symmetry and 
open/closed string duality at work.}
It is instructive to compare the above framework for $\cx N=1$ mirror
with the different approach of ref.\AVi.
In virtue of the genus zero duality \neosp\ this example will also 
illustrate the surprising one-to-one
mapping between closed string sphere \ffpi\ and open string disc
\wopexp\ instantons predicted by the genus zero part of the 
open/closed string duality conjecture.
We will only sketch the result for a specific 3-fold $Y$ here and 
refer to \wip\ more details on the computation, which 
involves a study of type IIA geometry and 
the phase structure of the open string moduli space.

We consider
a non-compact D-brane on the non-compact \CY $Y$ defined
as the canonical bundle $\cx O(-3)_{\IP^2}$.
We start from the linear sigma
model for the \CY 4-fold $X^*$ defined by eq. \defX, where
the charge vectors \defzt\ take the form
$$
I:\qquad \lm 0 = (1,-1,0,0,1,-1),\qquad \lm1=(-3,1,1,1,0,0)
$$
This describes the D5-brane with classical limit 
$\ty_0=z_0\, \ty_1 \ll \ty_1$.
From the differential equations \defGKZii, we obtain the
$\cx N=1$ mirror map \neomm\ and the superpotential \neosp.

By the relation \neosp, the superpotential $W$ describes at the same time the 
disc instanton corrections to the 4d $\cx N=1$ superpotential 
in the D-brane geometry $(Y,L)$ as well as
the sphere instanton corrections to the 2d superpotential 
on the dual 4-fold $X$. The open/closed string invariants 
are obtained from the integral large $t$ expansions \wopexp\ or
\ffpi\ respectively.
The result is displayed in Tab. 2\yyy for small degree.

\def\ss{\scriptstyle}
\vbox{
\vskip 0.1cm
\eqn\inst{
\vbox{\offinterlineskip\tabskip=0pt\halign{\strut
\vrule\hfil~$\ss{#}$~\hfil\vrule~
&\hfil$\ss{#}$~ &\hfil$\ss{#}$~ &\hfil$\ss{#}$~
&\hfil$\ss{#}$~ &\hfil$\ss{#}$~ &\hfil$\ss{#}$~
&\hfil$\ss{#}$~ &\hfil$\ss{#}$~ &\hfil$\ss{#}$~
\vrule\cr
\noalign{\hrule}
n_0&n_1=1&2&3&4&5&6&7&8&9\cr
\noalign{\hrule}
1& 2& -5& 32& -286& 3038& -35870& 454880& -6073311& 84302270\cr
2& 1& -4& 21& -180& 1885& -21952& 275481& -3650196& 50370000\cr
3& 1& -3& 18& -153& 1560& -17910&222588& -2926959& 40148496\cr
4& 1& -4& 20& -160& 1595& -17976& 220371& -2869120& 39055518\cr
\noalign{\hrule}
}}}
\noindent{\ninepoint  
{\bf Table 2\yyy}: 
Closed string invariants $D_{\vx k}$ of the toric variety $X$ in phase I. 
These agree with the open string invariants $d_{k,m}$ on $Y$, with the
vertical direction corresponding to the class of the basic disc
in $Y$ ending on $L$ and the horizontal directions to the class of 
the basic 2-sphere in $Y$.
}}

\ni
The above result is in full agreement with the 
calculation of the open string invariants $d_{k,m}$ in
the approach of ref.\AVi, see in particular Tab. 6 in ref.\AVii.

As $z_0$ grows, the perturbative expansion around the classical point 
$z_0=0$ breaks down near $z_0\sim1$.
The behavior of the superpotential
near $z_0=1$ can be studied from the solutions of the GKZ system
and it corresponds to a birational transformation, a ``flop'',
on the \CY 4-fold $X$ \wip.
Continuing to large $z_0\gg 1$ there is a new classical regime defined
by $y_1=z_0^{-1}y_0 \ll y_0$.  The appropriate
charge vectors $\lm \al$ that describe the perturbative expansion
in the new classical phase for $z_0\gg 1$ are defined by the
K\"ahler cone of $X$:
\eqn\moriII{
II:\qquad \lm 0 = (-1,1,0,0,-1,1),\qquad \lm1=(-2,0,1,1,1,-1)
}
and lead to the new invariants reported in Tab. 3\yyy.

\def\ss{\scriptstyle}
\vbox{
\vskip 0.1cm
\eqn\inst{
\vbox{\offinterlineskip\tabskip=0pt\halign{\strut
\vrule\hfil~$\ss{#}$~\hfil\vrule~
&\hfil$\ss{#}$~ &\hfil$\ss{#}$~ &\hfil$\ss{#}$~
&\hfil$\ss{#}$~ &\hfil$\ss{#}$~ &\hfil$\ss{#}$~
&\hfil$\ss{#}$~ &\hfil$\ss{#}$~ &\hfil$\ss{#}$~
\vrule\cr
\noalign{\hrule}
n_0&n_1=1&2&3&4&5&6&7&8&9\cr
\noalign{\hrule}
0&  1 &  -1 &  1 &  -2 &  5 &  -13 &  35 &  -100 &   300 \cr
1&  0 &  -2 &  4 &  -10 &  28 &  -84 &  264 &  -858 &     2860 \cr
2&  -1 &  0 &  12 &  -32 &  102 &  -344 &  1200 &  -4304 &      15730 \cr
3&  -1 &  5 &  0 &  -104 &  326 &  -1160 &  4360 &  -16854 &   66222 \cr
4&  -1 &  7 &  -40 &  0 &  1085 &  -3708 &  14274 &  -57760 &239404 \cr
5&  -1 &  9 &  -61 &  399 &  0 &  -12660 & 45722 &  -185988 &  793502\cr
6&-1 &  12 &  -93 &  648 &  -4524 &  0 &159208 &  -598088 &  2530946 \cr
7&  -1 &  15 &  -140 &  1070 &  -7661 &  55771 &0 &  -2112456 &  8171400 \cr
\noalign{\hrule}
}}}
\noindent{\ninepoint  
{\bf Table 3\yyy}: 
Closed string Gromov--Witten invariants of the toric variety $X$ in phase II. 
}}

\ni
The closed string invariants $D_{\vx k}$ in Tab. 3\yyy 
should be compared with the open string invariants $d_{k,m}$ of
the D-brane geometry specified by the above classical behavior.
Again the latter have been evaluated in the open string approach in
\AVii, see Tab. 5 therein. Again we find complete agreement
up to a trivial relabelling $n_0\to n_0-n_1$. The relabelling
corresponds to taking a linear combination of the charge vectors
$\lm \al$, which is fixed for us by the requirement that the 
$\lm \al$ span the dual of the K\"ahler cone of $X$. In particular
it is the choice \moriII\ which leads to positive intersection in the
phase II of $X$ - note that there is no such simple concept in 
the open string language where the same data would correspond
to the intersection calculus on bundles over $Y$.

In \AVii\ it was observed, that the non-perturbative superpotential
depends on the choice of an integral parameter, or ``frame'', that specifies 
the IR behavior of the D-brane\foot{See also the last reference in \vm\ 
for a further discussion.}. 
The above solution from $\cx N=1$ mirror symmetry does not have 
such a parameter, and defines a preferred frame. It would be interesting
to understand how the preferred frame is selected. It appears
that the fixing of the frame is linked to the expansion
around the point with maximal unipotent monodromy \wip.

\newsec{A local M-theory lift}
In the previous sections it has been proposed that the open 
D-brane geometries $(Y,L)$ are in some sense dual to
certain closed string \CY 4-fold compactifications 
without branes. It has also been verified that the
topological string amplitudes agree at $g=0,h=1$ and $g=0$,
respectively. Here we want to add some 
evidence that this coincidence might be the first term 
of an expansion of a true string theory duality between
the two backgrounds, using M-theory.

To start with, one may always take the large volume limit of $Y$
to reduce the type IIA geometry 
locally to a D6-brane on a SL cycle $L$ in the trivial $\CC^3$.
It has been argued in \AViii\ that this open string
background is dual to a closed string on the small resolution
of the conifold, by a combination of a lift to M-theory \AVii,
and the ``large $N$'' duality\foot{In fact $N=1$ in the present context.} 
between the open string on $T^*S^3$
and the conifold \vafaln\amv%
\ref\addtr{
J.~D.~Edelstein and C.~Nunez,
``D6 branes and M-theory geometrical transitions from gauged  supergravity,''
JHEP {\bf 0104}, 028 (2001), hep-th/0103167;\br
K.~Dasgupta, K.~Oh and R.~Tatar,
``Geometric transition, large N dualities and MQCD dynamics,''
Nucl.\ Phys.\ B {\bf 610}, 331 (2001), hep-th/0105066;\br
K.~Dasgupta, K.~Oh and R.~Tatar,
``Open/closed string dualities and Seiberg duality 
from geometric  transitions in M-theory,'', hep-th/0106040.}.
A necessary condition 
for a duality between the D-brane geometries
$(Y,L)$ and the \CY 4-fold $X$ is therefore that the latter must reduce
to the conifold in this limit. The 4-fold $X$ is the 
mirror of the manifold $X^*$ \defX, which has been proposed 
on the basis of its period structure in sect. 3\yyy. It is defined as 
a gauged linear sigma model with gauge group $U(1)^{h_{13}}$ 
and matter fields with $U(1)$ charges defined by the vectors $\lm \al$
in \defzt. One can verify that taking the image of the 
large volume limit of $Y$ in the moduli of $X$
leads to the local geometry 
$$
X\ \matrix{{\ss Vol(Y)\to \infty}\cr\longrightarrow\cr \phantom{x} } 
\ \cx O(-1)^{\oplus 2}_{\IP^1}\times \CC.
$$
This is precisely the small resolution of the conifold 
times two flat directions. Thus the proposed 
string duality for the geometries $(Y,L)$ and $X$ has passed
the first non-trivial test: in the local limit 
it reproduces the known string duality of \AViii.

One suspects that there is a generalization of the M-theory argument 
which extends to finite volume of $Y$ - 
and thus to non-trivial 3-folds. However the proof of \AViii\
relies also on the large $N$ duality for which there is no
known equivalent in the present case.
In the following we alter the argument for 
the duality of $\CC^3$ in a way that avoids the large $N$
duality and can be applied to the case of non-trivial
$Y$. 
 
Consider $\CC^3$, parametrized by $x_1,x_2,x_3$ and a
D-brane on a special Lagrangian 3-cycle $L$ defined by \AVi
$$
L:\qquad |x_1|^2-|x_3|^2=c_1,\quad |x_2|^2-|x_3|^2=c_2
$$
and $\sum \theta_i=\rmx{const}$, where $x_i=|x_i|e^{i\theta_i}$.
The 3-cycle $L$ has a boundary unless it ends on the locus $x_a=x_b=0$
for arbitrary $a,b$. Consider the phase with $c_2=0$, where the
brane ends on $x_2=x_3=0$. A holomorphic disc ending on $L$
is defined by 
$
D:\ x_2=x_3=0,\ |x_1|^2\leq c_1.
$
One may parametrize the radial direction of the 
disc by a real number $|x_1|^2=c_1-r_m$ and introduce
the M-theory circle as the phase of a coordinate $x_m=r_m^{1/2}e^{i\theta_m}$.
The $S^1$ fibration over $D$  defines the $S^3$ \AVii\AViii
\eqn\sthree{
S^3:\ \ x_2=x_3=0,\qquad |x_1|^2+|x_m|^2=c_1.
}
Moreover $r_m$ vanishes on $L$ and defines the D6-brane on $L$
in the M-theory reduction on the circle $x_m\to x_m e^{i\alpha}$. 
On the other hand, as
the only non-trivial cycle in this geometry is the $S^3$,
the local $G_2$ manifold for the M-theory must be the
spin bundle $\Sigma_{S^3}$ over $S^3$, 
which in turn allows another circle reduction 
to a type IIA string on the cotangent bundle $T^*S^3$. The
transition from the phase $c_1>0,c_2=0$ to a second phase with 
$c_1=0,c_2>0$ describe the large $N$ duality \vafaln\amv\
and establishes the equivalence of the two open string backgrounds to
the closed string on the small resolution $\cx O(-1)^{\oplus 2}_{\IP^1}$
of the conifold \AViii. The image of this transition 
under the moment map $x_i\to |x_i|^2$ is summarized in 
the figure below. The subscript IIA$_1$ refer 
to the geometry of the original type IIA theory with
the D6-brane on $L$ and the subscript IIA$_2$ to
the alternative $S^1$ reduction leading to a type IIA with D6-brane
on $T^*S^3$.

\def\smr{\cx O(-1)^{\oplus 2}_{\IP^1}}
{\goodbreak\midinsert
\centerline{\epsfxsize 5truein \epsfbox{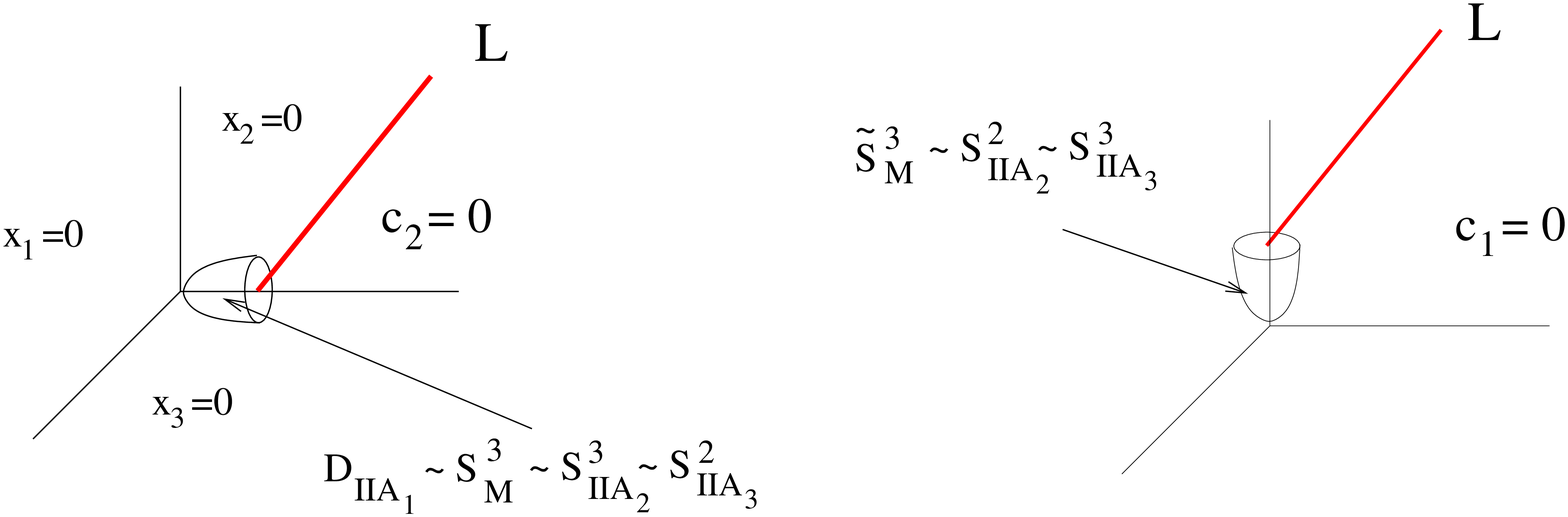}}
\leftskip 1pc\rightskip 1pc 
\noindent{\ninepoint  
{\bf Fig. 2\yyy}: The moment map for the D6-brane on $\CC^3$ with
the fundamental disc instanton $D$ for the two phases 
$c_2=0$ and $c_1=0$. The subscripts IIA$_n,\, n>1$ refer to the
two alternative $S^1$ reductions of the M-theory completion
described in the text. 
}\endinsert}
Since the three edges $x_a=x_b=0$ of $\CC^3$ are equivalent, it should be 
possible to avoid the large $N$ transition in the above argument
and to see the circle fibration that leads to type IIA on $\smr$ already 
in the original phase. 
The local $G_2$ manifold $\Sigma_{S^3}$, which is topologically
$\sim \bx R^4\times S^3$, may be described by the equation
\eqn\defS{
\Sigma_{S^3}:\ \ |x_1|^2+|x_m|^2-|x_2|^2-|x_3|^2=c_1,
}
where $x_2=x_3=0$ defines the $S^3$  and the $x_2,x_3$ parameterize 
$\bx R^4$. 

We are interested in a $U(1)$ acting on \defS\ without
fixed point which describes a third type IIA$_3$ compactification
in the phase $c_2=0$ (the result of a transition that starts
from a type IIA with D6-brane on $T^*S^3$ in the phase with $c_1=0$).
Specifically the $U(1)$ should 
act as the Hopf fibration $S^3\to S^2$ on the non-trivial $S^3$, 
defined by $(x_1,x_m)\to (e^{i\al}x_1,e^{i\al}x_m)$ in the above
coordinates. This is interpreted as the type IIA theory 
on the base $S^2$ with flux through the latter.

We make now the ansatz that the searched for M-theory $U(1)$ can be 
represented as a gauge symmetry of a 2d linear sigma model.
The motivation is that without the real constraint \defS,
the coordinates $x_1,x_2,x_3,x_m$ define a \CY 4-fold $\CC^4$.
Once we add the real constraint \defS\ to reduce to the local 
$G_2$ manifold, there is essentially one way to divide further
by a $U(1)$ to obtain a smooth \CY 3-fold with a 2-cycle. 
Namely we interpret \defS\ as the D-term $q_i\, |x_i|^2=c_1$ 
of the gauged $U(1)$. This determines in turn the $U(1)$ charges $q_i$
of the fields $x_i$. 
Note that the gauge transformation
$(e^{i\al}x_1,e^{i\al}x_m,e^{-i\al}x_2,e^{-i\al}x_3)$
with $\al\in\bx R$
acts precisely as the Hopf fibration on the $S^3$. It also
acts non-trivially in the $x_2,x_3$ directions, which may be
interpreted as flux in the directions transverse to the $S^2$ 
base.

The geometry of this $S^1$ reduction is precisely the same 
as that of the 
linear sigma model for the small resolution $\smr$. The momentum map
for the LSM with D-term \defS\ is shown in Fig. 3\yyy. It is 
in fact identical to the ``enlarged'' momentum map 
for the M-theory manifold described in \AViii.

A similar reasoning may now be applied to the case of D-branes
on a non-trivial 3-fold $Y$ with compact directions. For simplicity
we consider the example $Y=O(-3)_{P^2}$ described in sect. 4\yyy.
The argument applies more generally to other choices for $Y$ with some
obvious modifications.
Let $(x_0,x_1,x_2,x_3)$ be the coordinates on $Y$ with the 
projective action defined by the multiplication
$(\lambda^{-3}x_0,\lambda x_1,\lambda x_2,
\lambda x_3)$ where $\lambda\in \CC^*$. Let 
$L$ be the special Lagrangian 3-cycle in $Y$ defined by 
\eqn\defL{
|x_1|^2-|x_0|^2=c_1,\qquad |x_2|^2-|x_0|^2=c_2.
}
The D-term for the gauged $U(1)$ linear sigma model for $Y$ is 
\eqn\dtermii{
|x_1|^2+|x_2|^2+|x_3|^2-3|x_0|^2=t \ .
}
Consider the phase where the brane ends
on an interior edge, say $x_0=x_2=0$, which requires $c_2=0$. 
There are two holomorphic discs ending on $L$:
$$\eqalign{
D_1&:\ x_0=x_2=0,\quad 0\leq |x_1|^2\leq c_1,\cr
D_2&:\ x_0=x_2=0,\quad c_1\leq |x_1|^2\leq t \ .}
$$

To describe an M-theory limit we are looking for an $S^1$ fibration 
over $D_i$ that promotes an IIA instanton on the disc to a M2-brane wrapped
on $S^3$. For a single disc this can be achieved as in \sthree.
However a similar ansatz for the two discs is inconsistent with
the global structure constrained by the D-term $|x_1|^2+|x_3|^2=t$
on the edge $x_2=x_0=0$.

The best we can do is to introduce {\it two} new variables $x_4$ and $x_5$
that parametrize the radial direction of the two discs $D_i$ and 
define two 3-spheres located at $x_0=x_2=0$ and 
\eqn\tts{\eqalign{
S^3_1\ :\ x_4=0,\ &|x_1|^2+|x_5|^2=c_1+|x_4|^2,\cr
S^3_2\ :\ x_5=0,\ &|x_3|^2+|x_4|^2=t-c_1+|x_5|^2,}
}
These equations are now consistent with the D-term on the edge $x_0=x_2=0$.
Moreover the M-theory $S^1$ vanishing over $L$ is defined by 
the $U(1)$ action $(x_4,x_5)\to$$(e^{i\al}x_4,e^{i\al}x_5)$. 

Adding a term $|x_0|^2$ on the r.h.s. of the second equations 
in \tts, the local geometry is that of two spin bundles over 
the two homology 3-spheres, with the latter 
intersecting at the point $x_4=x_5=0$, and with the 
transverse directions identified in a non-trivial way. To make
these equations globally consistent with the D-term \dtermii\ we add
another term $|x_0|^2-|x_2|^2$ that is locally zero on the edge 
and obtain\foot{Adding the vanishing term instead to the other
equation is equivalent by a change of moduli.}
\eqn\mthst{\eqalign{
S^3_1\ :&\
|x_1|^2+|x_5|^2=c_1+|x_4|^2+|x_0|^2,\hskip90pt x_4=0,\cr
S^3_2\ :&\
|x_3|^2+|x_4|^2=t-c_1+|x_5|^2+|x_0|^2-(|x_2|^2-|x_0|^2),\
x_5=0.}
}
A new M-theory $U(1)$ without fixed points that defines a 
type IIA compactification without branes can now be 
obtained in a similar way as before. Namely, 
we interpret the equations \mthst\ as the D-terms of
a gauged linear sigma model with gauge symmetry $U(1)^2$.
This leads to the following $U(1)$ charges $\lm a_i$
for the  matter fields $x_i$:
\eqn\morilp{
\lm 1=(-1,1,0,0,-1,1),\qquad \lm 2=(-2,0,1,1,1,-1).
}
Note that the the $U(1)$ action defined by $\lm 1$ describes the 
Hopf fibration of the first $S^3$ and the $U(1)$ 
$\lm 2$ the Hopf fibration of the second $S^3$ in \mthst. 
Moreover the diagonal $U(1)$ coincides with the gauge symmetry
of the GLSM for $Y$. 

\def\smr{\cx O(-1)^{\oplus 2}_{\IP^1}}
{\goodbreak\midinsert
\centerline{\epsfxsize 5truein \epsfbox{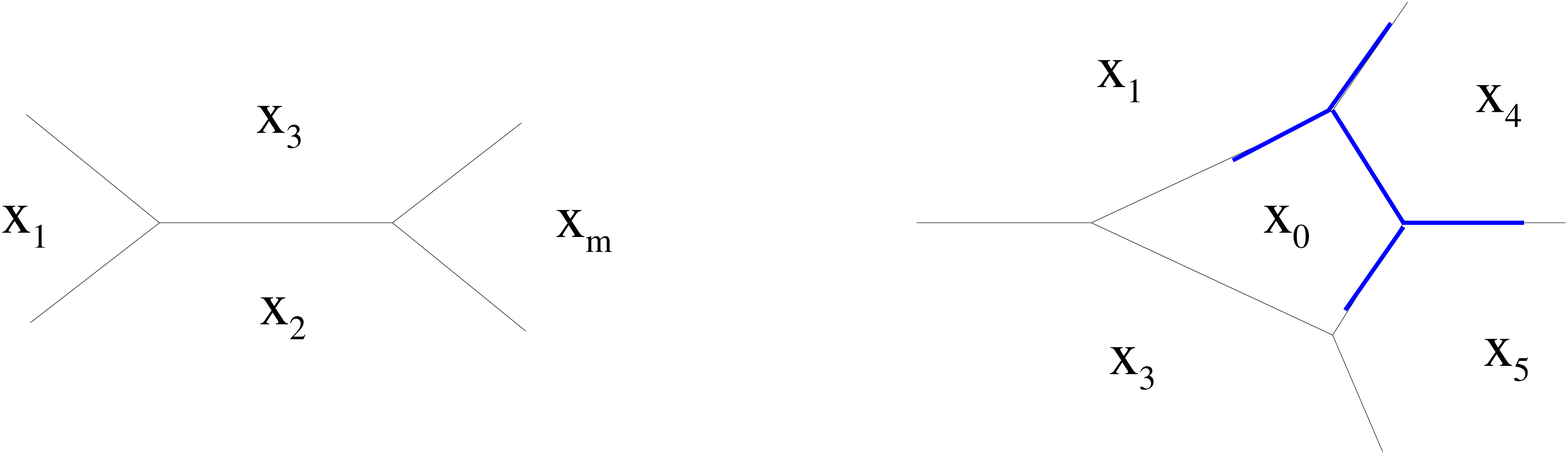}}
\leftskip 1pc\rightskip 1pc 
\noindent{\ninepoint  
{\bf Fig. 3\yyy}: The moment map for the LSM defined by 
\defS\ and the projection to $x_2=0$ of the momentum map
for the LSM on the 4-fold \mthst. 
We have also indicated how the momentum map for the spin bundle 
$\Sigma_{S_1^3}$ sits in the latter (blue lines). 
}\endinsert}

The geometry for the GLSM defined above is precisely the mirror 
of the \CY 4-fold $X^*$ defined in \defX! 
Moreover the two Hopf fibrations of the M-theory circle 
over the $S^2$ bases indicates the presence of flux on the 4-fold.
The flux generates (part of) the superpotential 
in the two-dimensional type IIA theory on $X$.

Note that to realize the M-theory lift of the discs to 3-spheres in 
\tts, we needed to add two complex variables $x_4$ and $x_5$.
Subtracting the one real constraint this adds three real 
dimensions to the six of the \CY 3-fold $Y$. This is in agreement 
with the 2d interpretation of the duality in sect. 2\yyy.

In the above argument we have clearly changed the global 
geometry transverse to the edge $x_2=x_0=0$ on which the discs and their
M-theory lifts to $S^3$ are defined. This is not unexpected, as the
closed string includes the back-reaction of the geometry 
to the D-branes; yet it makes this reasoning somewhat heuristic.
Note however that the same is already true for the M-theory lift for
$\CC^3$ \sthree,
where the string duality has been tested at all genera. One may
therefore hope that this {\it local} M-theory lift captures 
the essential geometry and the transverse geometry is uniquely fixed 
by the consistency of the background. Clearly it would be very
satisfying to have an independent check of this type of arguments
by comparing amplitudes other then the superpotential.

\newsec{Outlook}
There are several interesting questions arising from the 
previously described connection between the exact 4d $\cx N=1$
superpotential and the 2d topological closed string amplitudes.
Firstly one would like to generalize these ideas to 
more general D-brane geometries, involving many D-branes
and compact \CY 3-folds $Y$. Also the generalization to
higher genera will be interesting, both from the math/physics
point of view as well as to test the conjectured duality. 
On more conceptual grounds, a quite interesting structure
is the $\cx N=1$ 4d special geometry inherited from the 
2d amplitudes. It would be interesting to derive it
from the open string equivalent of the $tt^*$ equations. 
These questions will be addressed in \wip.

Another interesting direction is 
F-theory on the 4-fold as a natural candidate to lift up the 
conjectural duality to four dimensions\foot{An F-theory argument
in favor of the proposed open/closed string duality has been pointed out
to me by Cumrun Vafa.}. 
In fact the same 2d topological amplitudes that we have
considered here have been proposed to be relevant for
F-theory in the earlier study of mirror symmetry for
4-folds \ffsp. Our result indicates that this is indeed
the case. It would be interesting to have an alternative 
2d world-sheet formulation - describing F-theory -
that leads to the same amplitudes as that of the closed
type IIA string considered in this paper.

\vskip10pt\ni
{\bf Acknowledgments:}\br
I am grateful to Mina Aganagic, Wolfgang Lerche
and Cumrun Vafa for valuable discussions and explanations.

\appendix{A}{Geometric integrals for the dual open/closed string pairs}
In the following we show that the period integrals on the closed
string background $X^*$ without branes reproduce the 
3-cycle and 3-chain integrals of the proposed dual open string background 
$(C,Y^*)$. The Landau--Ginzburg expression of the 
period integrals over the holomorphic $d$-form $\OM d$ is \cec
\eqn\lgper{\eqalign{
\Pi&\sim\int \prod_0^{d-1} \fc{dy_i}{y_i}\ e^{-\Wtd(y_i)}\cr
&=\int \prod_0^{d-2} \fc{d\tx y_i}{\tx y_i}dy_{d-1}\, dx\, dz\ 
e^{-y_{d-1}(\Wtd(\tx y_i)+xz)},\cr
&=\int_S \prod_0^{d-2} \fc{d\tx y_i}{\tx y_i}\fc{dz}{z},
\hskip 35pt \tx y_i=\fc{y_i}{y_{d-1}},\ i=1,...,d-2,}}
where in the last, alternative expression the integral is on the hypersurface 
\eqn\hypo{
S:\, \Wtd(\tx y_i)+xz=0,}
with $x,z \in \CC$ \HIV. The equivalence of the
above representations follows readily from the integration
of linear variables, $\int dx e^{xf}=\delta(f)$.
In particular integrating out $x,z$ in the second line gives back the
first one, while integrating out $y_{d-1}$ leads to the expression in the 
third line. We have also used the homogeneity of $\Wtd$ in the variables
$y_i$ which assures that $y_{d-1}$ factors out of $\Wtd$ in the variables 
$\tx y_i$.

To show that periods of the holomorphic form $\OM4(X^*)$ on the 
closed string geometry $X^*$ \defX\ describes the 3-cycle and 3-chain integrals
over $\OM3$ in the D-brane geometry $(Y^*,C)$, 
consider the integrals 
\eqn\genXper{
\Pi\sim\int \prod_0^{2} \fc{dy_i}{y_i}\fc{dv}{v}\ 
e^{-\Wtd(Y^*)-v\,(1+\zo ^{-1}\fc{y_0}{y_1})},}
where we have solved for $v_2$ and write $v$ for $v_1$
for ease of notation. Note that the variable $v$ enters only 
{\it linearly} into $\Wtd(X^*)$. 

As explained in sect. 3\yyy, there are two types of 
4-cycles in the manifold $X^*$, considered as a fibration of
$Y^*$ over an infinitely long cylinder parametrized
by $v$.
The period integrals for the first type of cycles is 
obtained by integrating on a small cycle around $v=0$ and one finds
\eqn\genXperii{
\Pi\sim\int \prod_0^{2} \fc{dy_i}{y_i}\ e^{-\Wtd(Y^*)}.}
These are simply the periods of $Y^*$, times an irrelevant
constant. This is also obvious from the fibration.

The second type of 4-cycles projects onto a path between
branch points on $\CC^*_v$, above which a 3-cycle shrinks
in the fiber $Y^*$.
To evaluate this type of integral, we 
take a derivative with respect to $\tau_0=\fc{1}{2\pi i}\ln \zo $:
$$
\eqalign{
\fc{\p}{\p \tau_0}\Pi &\sim 
\, \zo ^{-1}
\int\prod_{i=0}^2\fc{dy_i}{y_i}dv\fc{y_0}{y_1}
\ e^{-\Wtd(Y^*)-v\,(1+\zo ^{-1}\fc{y_0}{y_1}) }\cr
&=\, \zo ^{-1}
\int\prod_{i=0}^2\fc{dy_i}{y_i}\fc{y_0}{y_1}\, 
\delta(1+\zo ^{-1}\, \fc{y_0}{y_1})\, e^{-\Wtd(Y^*)}.}
$$
Note that left-over integral in the last expression is again 
defined on the \CY 3-fold $Y^*$, however this time with an extra delta function
produced by the integration over the linear variable $v$.

To proceed we assume small $z_0$ and integrate out $y_0$. Moreover
we use the simple manipulation described below \lgper\ to obtain 
the equivalent expression
$$
\fc{\p}{\p \tau_0}\Pi \sim \int_{S} \fc{d\tx y_1}{\tx y_1}\fc{dz}{z},
$$
where $S$ is the surface 
$$
S:\ \Wtd(Y^*)|_{\tx y_0=\zo  \tx y_1}+xz=0,
$$
\ni
The surface $S$ has the structure of a $\CC^*$ fibration over the 
$\tx y_1$ plane and there are again two types of 2-cycles made from a 1-cycle 
in the $\CC^*$ fiber and a path in the base. 
Integration on small circles around $z=\tx y_1=0$ gives a constant.
This is the expected result for the period that describes the 
flat coordinate 
$$
t_0=\tau_0+ S_0(z_\al),\qquad \al=0,...,h_{13}-1.
$$
where $S_0$ is the the sub-leading correction to the
period integral discussed in sect.4\yyy. However note that 
the derivative 
$\p/\p{\tau_0}=\sum_{a>0}(\p t_a/\p\tau_0)\p/\p{t_a}
+(\p t_0/\p\tau_0)\p/\p{t_0}$
is in general different from $\p/\p{t_0}$. The above outcome suggests that 
$\p t_a/\p\tau_0=0$ for $a>0$ and $\p t_0/\p\tau=1$, which implies that the
sub-leading corrections $S(z_\al)$ 
to the flat coordinates are functions of {\it only} the 
$h_{13}-1$ coordinates $z_a,\, a>0$. It is not too 
difficult to verify this relation by showing that 
the periods that satisfy $\p\Pi/\p t_a=1,\, a>1$ are independent of $\zo $. 
A more powerful approach is to consider the system \defGKZii\ of 
differential equations satisfied by the periods \genXper, and 
it has been already shown in sect. 4\yyy that this leads to the desired 
result $\p/\p\tau_0=\p/\p t_0$.

There is another 2-cycle in the surface $S$ made from 
integrating $\tx y_1$ instead from infinity to the special points $\tx y_1^0$,
where the cycle in the $\CC^*$ fiber shrinks. The result for
this type of integral is 
$$
\fc{\p}{\p t_0}\Pi \sim 
\int_\infty^{\tx y_1^0} d(\ln(y_1))=\ln(y^0_1)+\dots,
$$
where the dots denote the contribution from infinity which
is independent of the moduli. This is precisely the 
first derivative of the 
superpotential \spavi\ for the D5-brane on $Y^*$.

\listrefs
\end